\documentclass{aa}

\usepackage[varg]{txfonts}

\usepackage{verbatim}
\usepackage{graphicx}
\usepackage{epstopdf}
\usepackage{subfigure}
\usepackage{hyperref}
\usepackage{mathtools}
\usepackage{mathrsfs}
\usepackage{amsmath}

\title{On the Fast Magnetic Rotator Regime of Stellar Winds}

\titlerunning{On the Fast Magnetic Rotator Regime of Stellar Winds}

\author{C. P. Johnstone\inst{\ref{vienna}}}

\institute{
University of Vienna, Department of Astrophysics, T\"{u}rkenschanzstrasse 17, 1180 Vienna, Austria \label{vienna}
}

\abstract{}{
We study the acceleration of the stellar winds of rapidly rotating low mass stars and the transition between the slow magnetic rotator and fast magnetic rotator regimes. 
We aim to understand the properties of stellar winds in the fast magnetic rotator regime and the effects of magneto-centrifugal forces on wind speeds and mass loss rates.
}{
We extend the solar wind model of \citet{Paper1} to 1D magnetohydrodynamic (MHD) simulations of the winds of rotating stars.
We test two assumptions for how to scale the wind temperature to other stars and assume the mass loss rate scales as $\dot{M}_\star \propto R_\star^2 \Omega_\star^{1.33} M_\star^{-3.36}$, in the unsaturated regime, as estimated by \citet{Paper2}.
}{
For 1.0~M$_\odot$ stars, the winds can be accelerated to several thousand km~s$^{-1}$, and the effects of magneto-centrifugal forces are much weaker for lower mass stars.
We find that the different assumptions for how to scale the wind temperature to other stars lead to significantly different mass loss rates for the rapid rotators. 
If we assume a constant temperature, the mass loss rates of solar mass stars do not saturate at rapid rotation, which we show to be inconsistent with observed rotational evolution. 
If we assume the wind temperatures scale positively with rotation, the mass loss rates are only influenced significantly at rotation rates above $\sim 75\Omega_\odot$.
We suggest that models with increasing wind speed for more rapid rotators are preferable to those that assume a constant wind speed.
If this conclusion is confirmed by more sophisticated wind modelling. it might provide an interesting observational constraint on the properties of stellar winds.
}{}

\begin{document}

\maketitle


\section{Introduction}


The solar wind accelerates because of a combination of thermal pressure gradients and pressure from waves propagating through the wind (\citealt{Cranmer07}; \citealt{Cranmer09}).
For the winds of rapid rotators, a further mechanism is important.
Due to the fast rotation and the stellar magnetic field, the wind is accelerated by magneto-centrifugal forces (\citealt{BelcherMacGregor76}). 
For the most rapid rotators, the winds can reach speeds of several thousand km~s$^{-1}$. 
The rotation rates at which magneto-centrifugal forces are important are determined by the stellar magnetic field strength and the contributions of the other acceleration mechanisms (\citealt{HolzwarthJardine07}). 

All these effects are important given the influence of stellar winds on the atmospheric evolution of planets. 
Stellar winds can cause planetary atmospheres to lose mass (\citealt{Kislyakova13}) and can potentially change atmospheric chemistry and even the surface climate (\citealt{Airapetian16}).
At young ages, low-mass stars show a large spread in rotation rates which quickly converge as stars spin down (\citealt{GalletBouvier13}).
A subset of stars go through phases of extremely rapid rotation, with rotation rates exceeding 100 times that of the Sun (\citealt{Hartman10}).
The magnetic activity of such stars evolve significantly differently from their slowly rotating counterparts (\citealt{Paper2}; \citealt{Tu15}) which can be important for the evolution of a planet's atmosphere (\citealt{Johnstone15}).

The basic mathematical theory for the equatorial winds of rotating magnetised stars was developed by \citet{WeberDavis67}.
\citet{BelcherMacGregor76} defined two regimes: the slow magnetic rotator (SMR) regime and the fast magnetic rotator (FMR) regime.
In the SMR regime, magneto-centrifugal forces are negligible; in the FMR regime, they can be the dominant process determining the velocity of the wind.
Extremely rapid rotators can enter the centrifugal magnetic rotator (CMR) regime when the equatorial corotation radius, $R_\mathrm{co}$, is in the subsonic part of the wind; in the CMR regime, the mass flux in the wind is massively enhanced by magneto-centrifugal effects.
When $R_\mathrm{co}$ is smaller than the stellar radius, the star breaks apart. 

The aim of this paper is similar to those of \citet{BelcherMacGregor76}, though we use improved understandings of the solar wind, the magnetic fields of Sun-like stars, wind mass loss rates, and stellar rotational evolution; we also apply a 1D MHD solar wind model.  
In Section~\ref{sect:model}, we describe our wind model; in Section~\ref{sect:ModelAorB}, we consider the different assumptions about wind temperature on the mass loss and spin-down of rapid rotators; in Section~\ref{sect:FMRwinds}, we study the properties of winds in the FMR regime and the transition between the SMR and FMR regimes; in Section~\ref{sect:conclusions}, we discuss our results\footnotemark.

\footnotetext{All of the codes and output data used in this paper can be downloaded from \url{https://goo.gl/hTuEVw} or obtained by contacting the author.}


\section{Wind model} \label{sect:model}

\begin{figure}
\centering
\includegraphics[trim = 0mm 0mm 0mm 0mm, clip=true,width=0.45\textwidth]{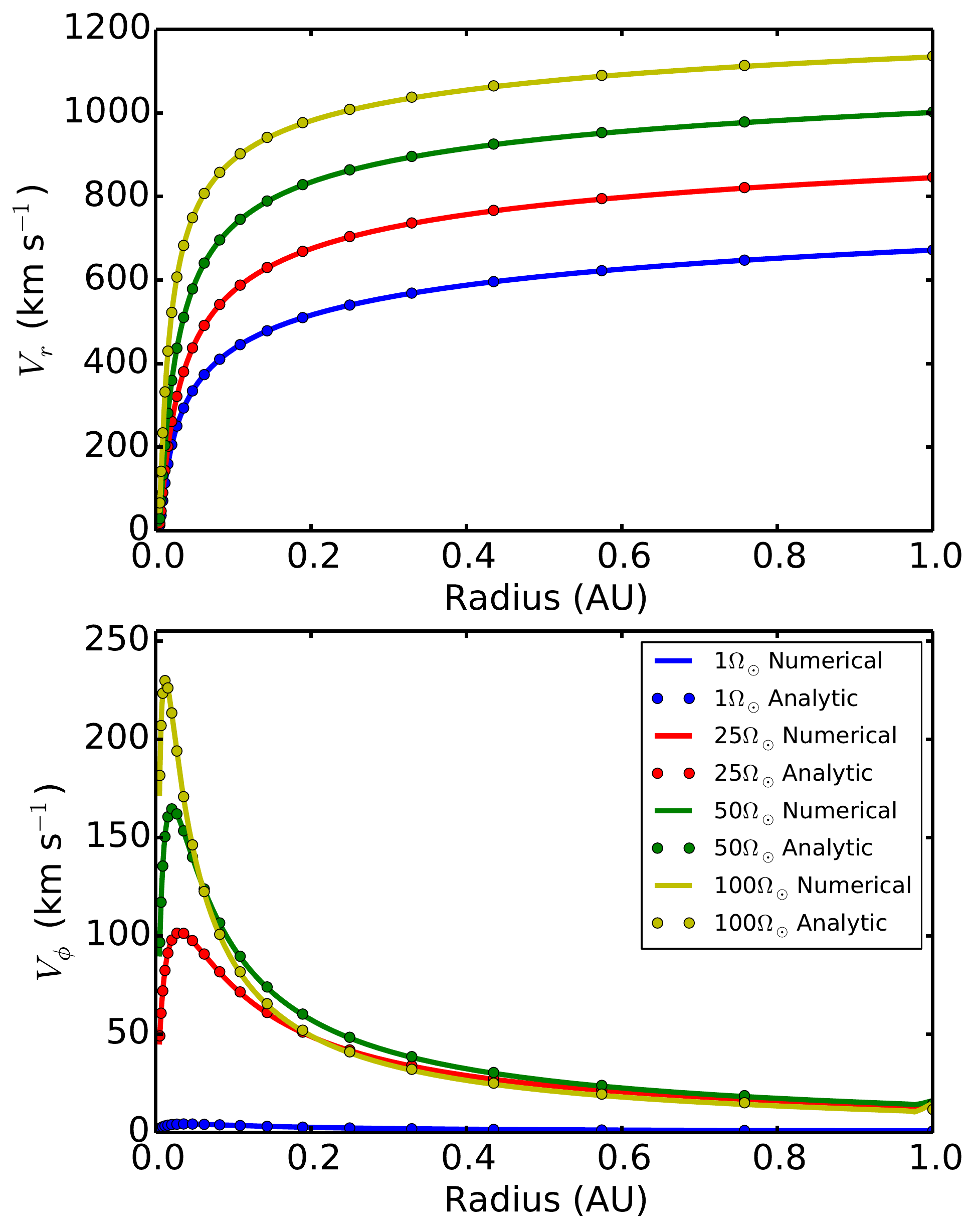}
\caption{
Figure comparing the velocity structures of isothermal winds simulated using our numerical MHD model (solid lines) and using the analytic Weber-Davis model (circles) for different rotation rates.}
 \label{fig:VisoComparison}
\end{figure}

\subsection{MHD code}

Our 1D MHD wind model is based on the hydrodynamic model developed by \citet{Paper1} and is described in detail in their Section~3.1 and Appendix~A.
The model is solved using the \emph{Versatile Advection Code} (VAC) developed by \citet{Toth96}.
It is common in stellar wind models to assume that the thermal pressure, $p$, and density, $\rho$, are related by a polytropic equation of state, such that \mbox{$p = K \rho^\alpha$}.
In our model, we take \mbox{$\alpha = 1.05$} when \mbox{$r < 15 R_\odot$} and \mbox{$\alpha = 1.51$} when \mbox{$r < 25 R_\odot$}.
Between 15 and 25 $R_\odot$, we assume $\alpha$ varies between these values linearly. 
In \citet{Paper1}, the radial structures of our model were verified against measured values of the real solar wind, both close to the Sun and far away.

In this paper, we extend the model to take into account magnetic fields and stellar rotation.  
We therefore use the MHD physics model in VAC, modified to take into account the spatial variations in $\alpha$.
In addition, we include momentum source terms for centrifugal and Coriolis forces. 
We hold $B_r$ constant throughout the simulation domain, as is necessary in 1D MHD.   
At the lower boundary, we hold the density, $\rho$, constant and allow $B_\phi$ to vary with no constraints; the radial velocity component, $V_r$, in the first grid point is set to be equal to the value in the second grid point and $V_\phi$ at the lower boundary is set such that the velocity and magnetic field vectors are parallel. 
At the upper boundary, we assume zero spatial gradients in all quantities.

\begin{figure}
\centering
\includegraphics[trim = 0mm 0mm 0mm 0mm, clip=true,width=0.49\textwidth]{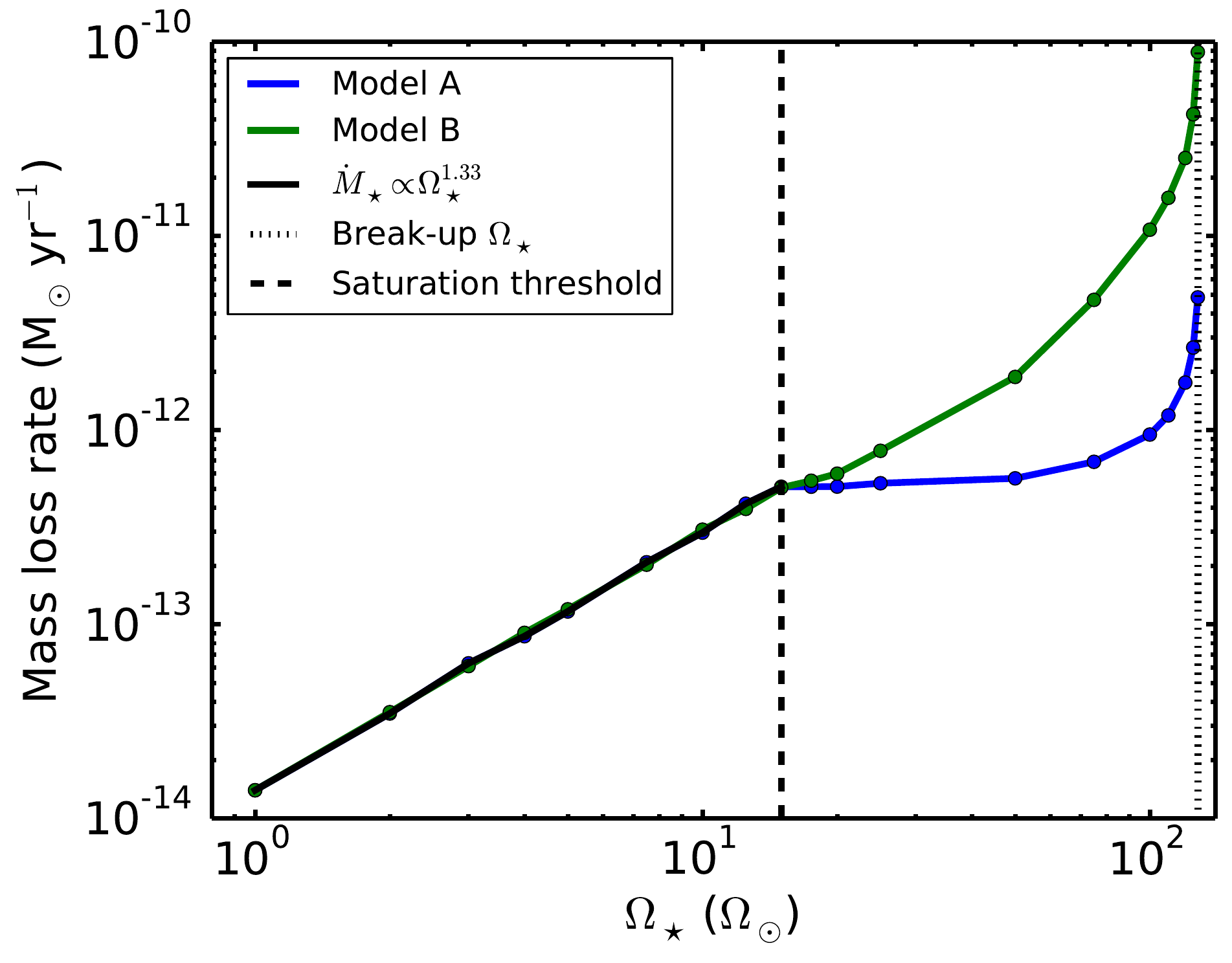}
\caption{
Figure showing wind mass loss rate against rotation rate for both Model~A and Model~B for solar mass stars. 
In both cases, $\dot{M}_\star$ follows Eqn.~\ref{eqn:Mdot} in the unsaturated regime and is then allowed to vary freely in the saturated regime. 
The filled circles show individual simulations.
}
\label{fig:MdotsSolar}
\end{figure}

We start the simulations with a simple isothermal Parker wind solution (\citealt{Parker58}) for $V_r$ at all radii.
The initial density structure is calculated based on mass conservation (i.e. \mbox{$\rho V_r r^2 = \mathrm{constant}$}) and the radial magnetic field is calculated from \mbox{$B_r (r) = B_\star \left(r/R_\star \right)^{-2}$}, where $B_\star$ is the assumed radial magnetic field at the stellar surface.
Both $V_\phi$ and $B_\phi$ are zero everywhere in the initial conditions.
Early in the simulation, a non-physical shock propagates through the domain and the simulation quickly relaxes to its final state.  
The simulations are performed on a grid of 500 cells extending from the stellar surface to 1~AU.
The sizes of each grid cell increase linearly from the stellar surface to the outer boundary, with the final grid point being 100 times larger than the first.

\subsection{Validation of code}

To verify our simulations, we run models for isothermal winds (i.e. \mbox{$\alpha=1$} everywhere) with different rotation rates and compare the results to solutions of the analytic model of \citet{WeberDavis67}. 
The method we use to solve the Weber-Davis (WD) model is described in Appendix~\ref{appendix:analyticWD}. 

For the MHD simulations, we assume at the base of the wind (i.e. at the stellar surface), a proton number density of $10^7$~cm$^{-3}$, a wind temperature of 2~MK, and a  radial magnetic field of 1~G.  
We run simulations with stellar rotation rates of $1 \Omega_\odot$, $25 \Omega_\odot$, $50 \Omega_\odot$, and $100 \Omega_\odot$. 
For the analytic models, the only difference is that we use the mass loss rates from the MHD simulations as an input instead of the base density.
This is necessary because the analytic model requires the mass flux as an input. 
In Fig.~\ref{fig:VisoComparison}, we compare the radial structures of $V_r$ and $V_\phi$ from both methods and find excellent agreements.
There are similarly good agreements in other wind parameters, and also for simulations with stronger magnetic fields.

\begin{figure}
\centering
\includegraphics[trim = 0mm 0mm 0mm 0mm, clip=true,width=0.45\textwidth]{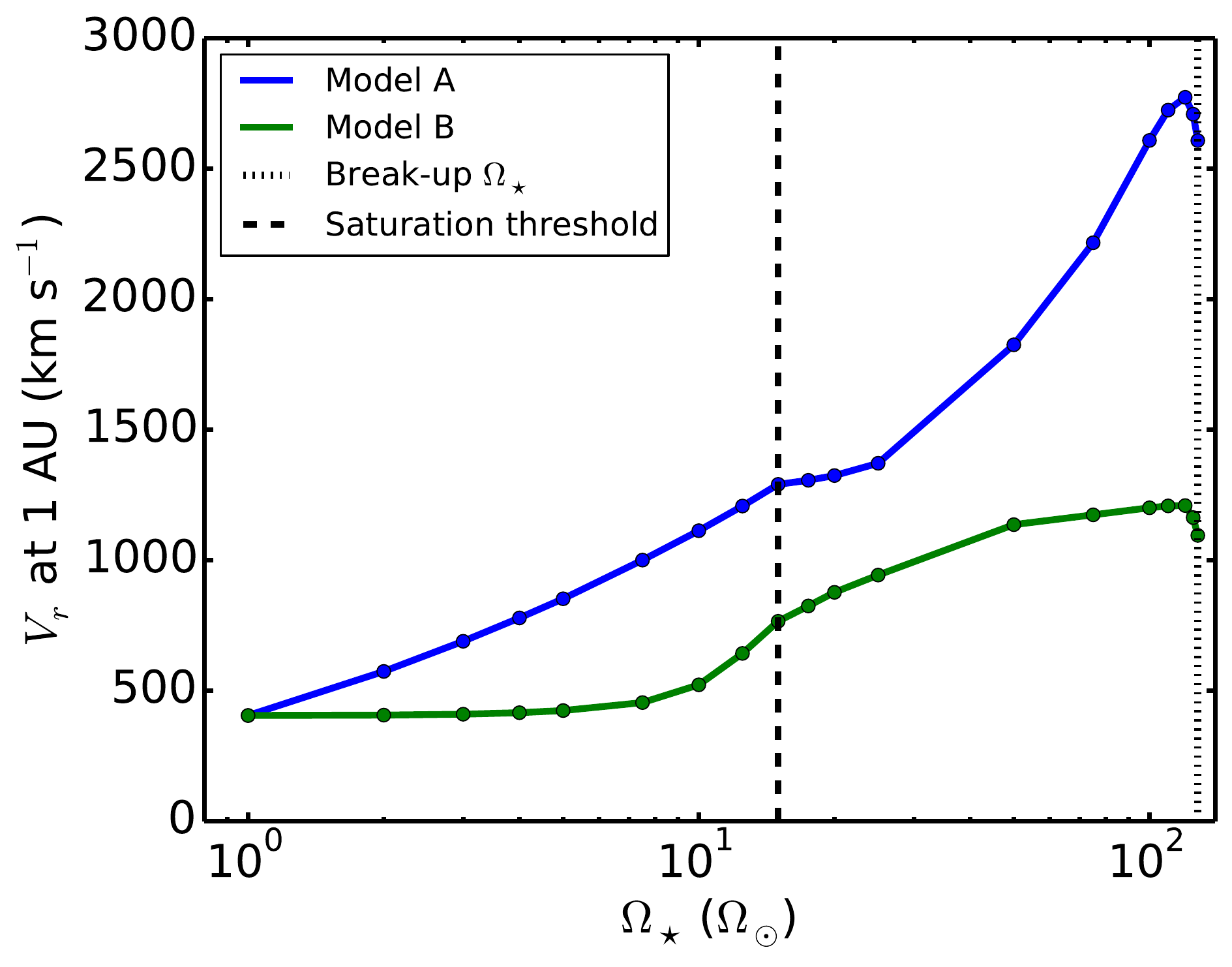}
\includegraphics[trim = 0mm 0mm 0mm 0mm, clip=true,width=0.45\textwidth]{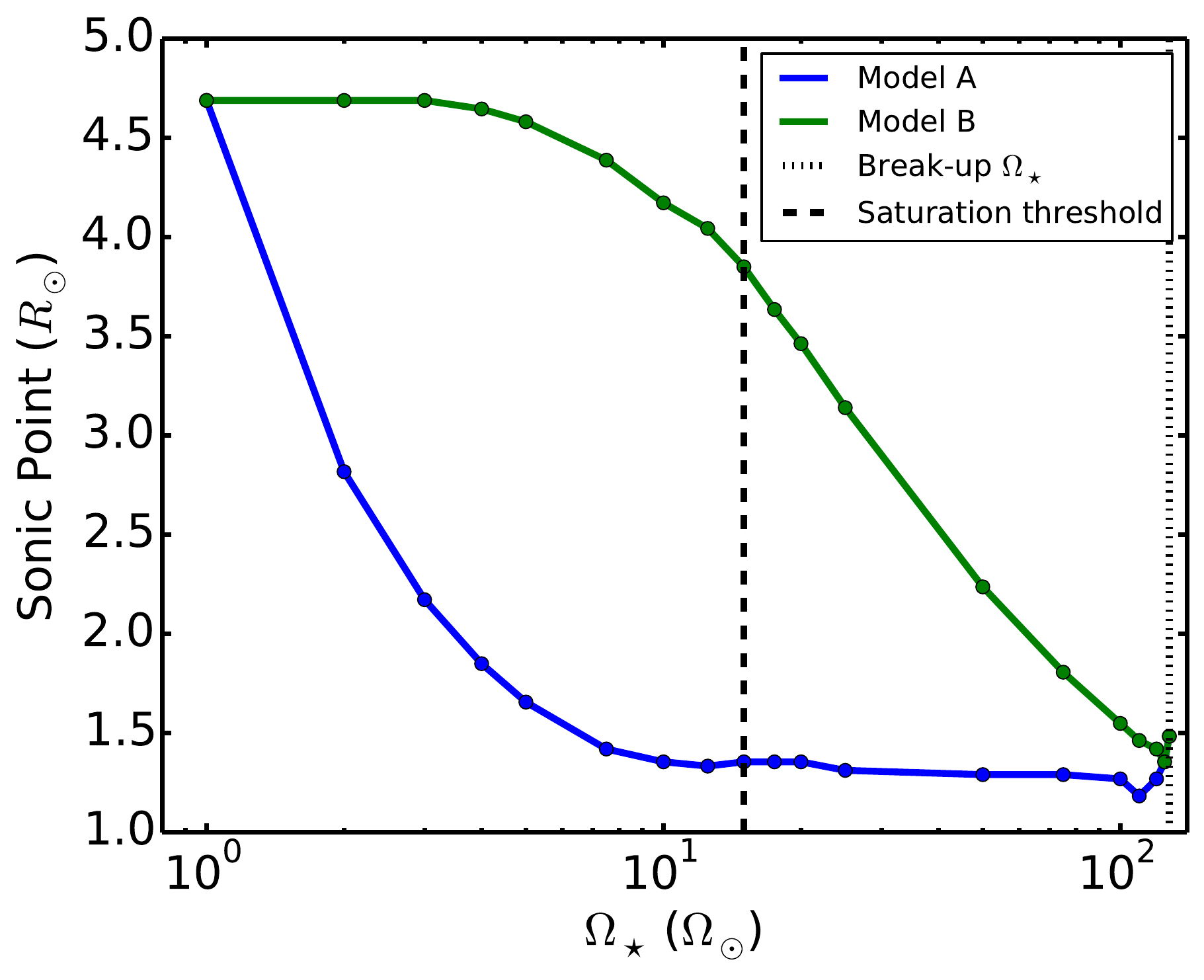}
\caption{
Figure showing the rotational dependence of the wind speed at 1~AU (\emph{upper-panel}) and the radius of the sonic point (\emph{lower-panel}) for Model~A and Model~B. 
The filled circles show individual simulations.
}
\label{fig:SonicSolar}
\end{figure}

\subsection{Application to stars}

\begin{figure}
\centering
\includegraphics[trim = 0mm 0mm 0mm 0mm, clip=true,width=0.45\textwidth]{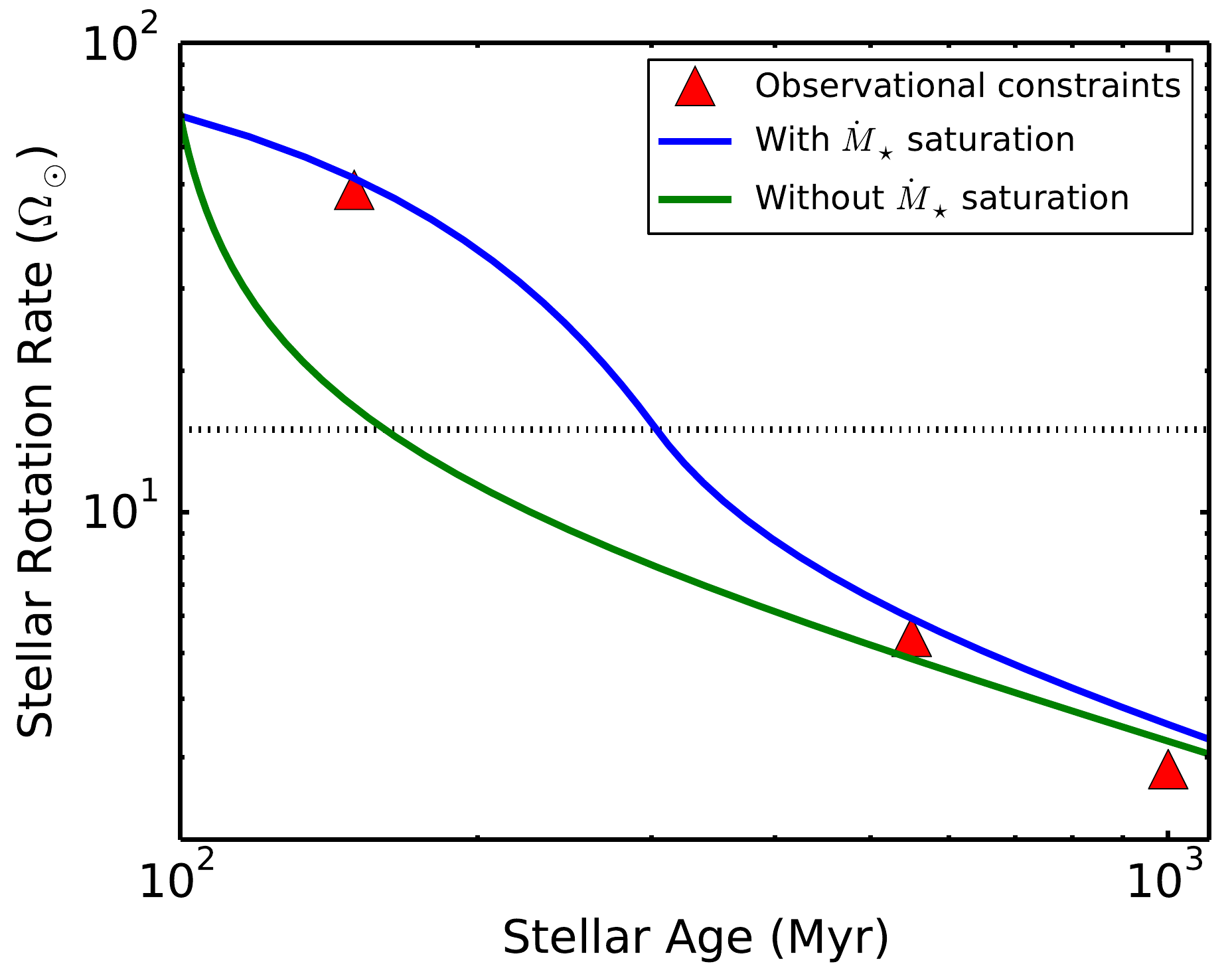}
\caption{
Figure showing the rotational evolution between 100~Myr and 1~Gyr of a solar mass star with a rotation rate at 100~Myr of $70\Omega_\odot$.
The blue and green lines show the evolutions with and without sautation of $\dot{M}_\star$ at $15\Omega_\odot$.
These correspond approximately to the $\dot{M}_\star$--$\Omega_\star$ relations for Model~A and Model~B shown in Fig.~\ref{fig:MdotsSolar}.
}
\label{fig:RotEvo}
\end{figure}

We scale our solar wind model to other stars based on their masses and rotation rates.
This is similar to the approach taken by \citet{HolzwarthJardine07}, and more recently using 3D MHD models by \citet{Reville16}.
For the radius and luminosity, we simply assume \mbox{$R_\star \propto M_\star^{0.8}$} and \mbox{$L_\star \propto M_\star^{3.9}$}.
Since we are interested in winds in the equatorial plane, we consider only the slow component of the solar wind.
The three parameters in our model are the temperature, density, and magnetic field strength at the base of the wind.   
For the current solar wind, these were determined by \citet{Paper1} to be 1.8~MK and \mbox{$2.6 \times 10^{7}$~cm$^{-3}$} for the temperature and proton density respectively. 
We assume a base magnetic field of 0.54~G, as justified in Section~4.3 of \citet{Paper1}; this value takes into account the fact that we assume \mbox{$B_r \propto r^{-2}$} everywhere, whereas in reality, $B_r$ decreases faster with $r$ within the closed corona. 

We assume that the wind temperature scales linearly with coronal temperature, which is known to depend strongly on X-ray activity (\citealt{Schmitt97}; \citealt{Gudel97}).
We use the relation derived by \citet{JohnstoneGudel15} for the coronal average temperature, $\bar{T}_\mathrm{cor}$, in MK as a function of X-ray surface flux $F_\mathrm{X}$, in erg~s$^{-1}$~cm$^{-2}$:
\begin{equation}
\bar{T}_\mathrm{cor} = 0.11 F_\mathrm{X}^{0.26}.
\end{equation} 
To get $F_\mathrm{X}$ as a function of stellar mass and rotation rate, we use the empirical relation derived by \citet{Wright11}.
The result is that winds are hotter for more rapidly rotating stars until $F_\mathrm{X}$, and therefore $\bar{T}_\mathrm{cor}$, saturates.
Saturation takes place at lower $\Omega_\star$ for lower-mass stars. 
The above model was called Model~A by \citet{Paper1}.
In the same paper, \citet{Paper1} considered a separate assumption, which they called Model~B.
In this model, they assumed, as in \citet{Matt12}, that the sound speed at the base of the wind is a constant fraction of the surface escape velocity.
The wind temperature in this model is approximately the same for all stars. 
\citet{Paper1} showed that these two models can lead to very different wind properties. 
In Section~\ref{sect:ModelAorB}, we test both models and describe why we prefer Model~A to Model~B.

\begin{figure*}
\centering
\includegraphics[trim = 0mm 0mm 0mm 0mm, clip=true,width=0.45\textwidth]{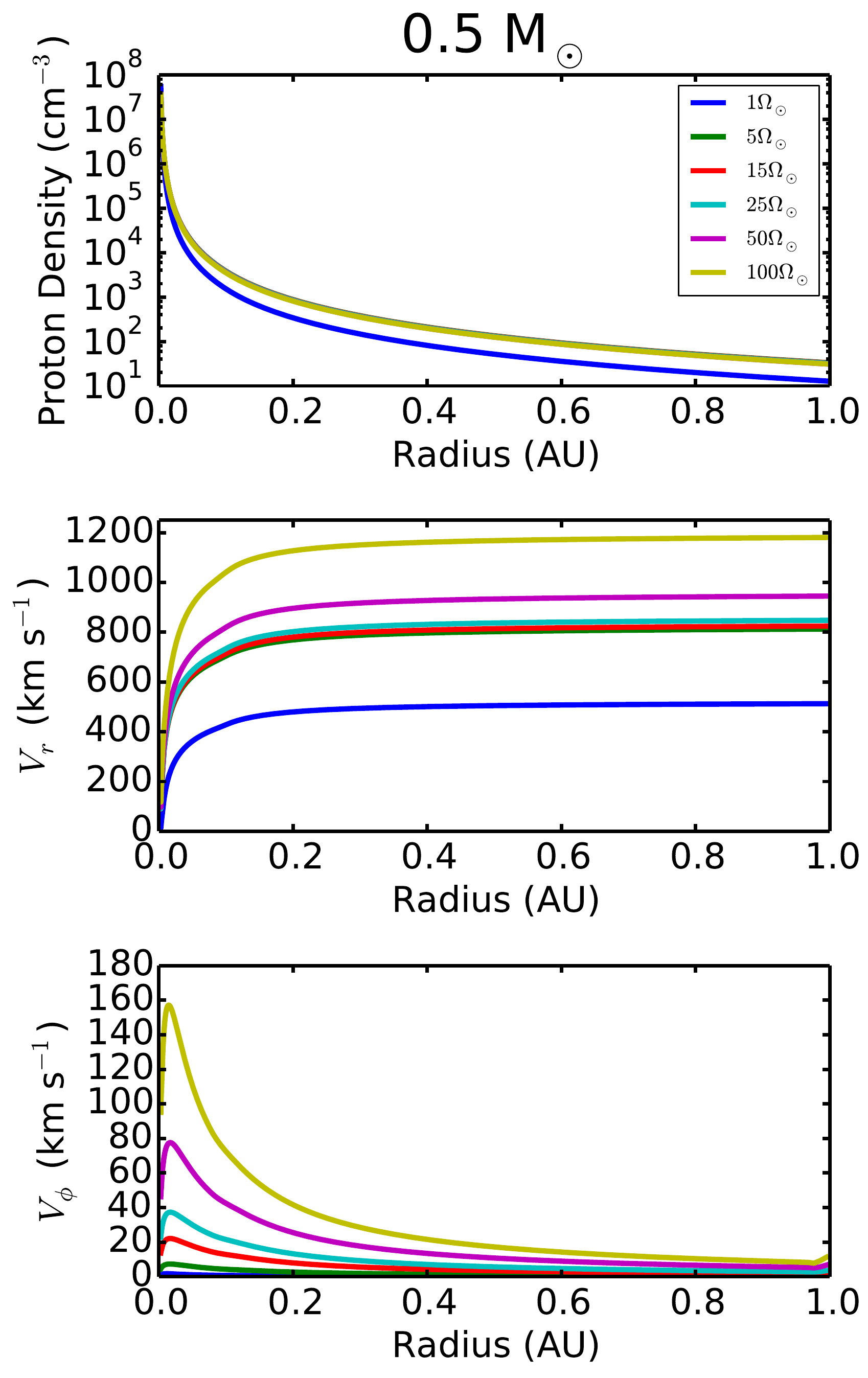}
\includegraphics[trim = 0mm 0mm 0mm 0mm, clip=true,width=0.45\textwidth]{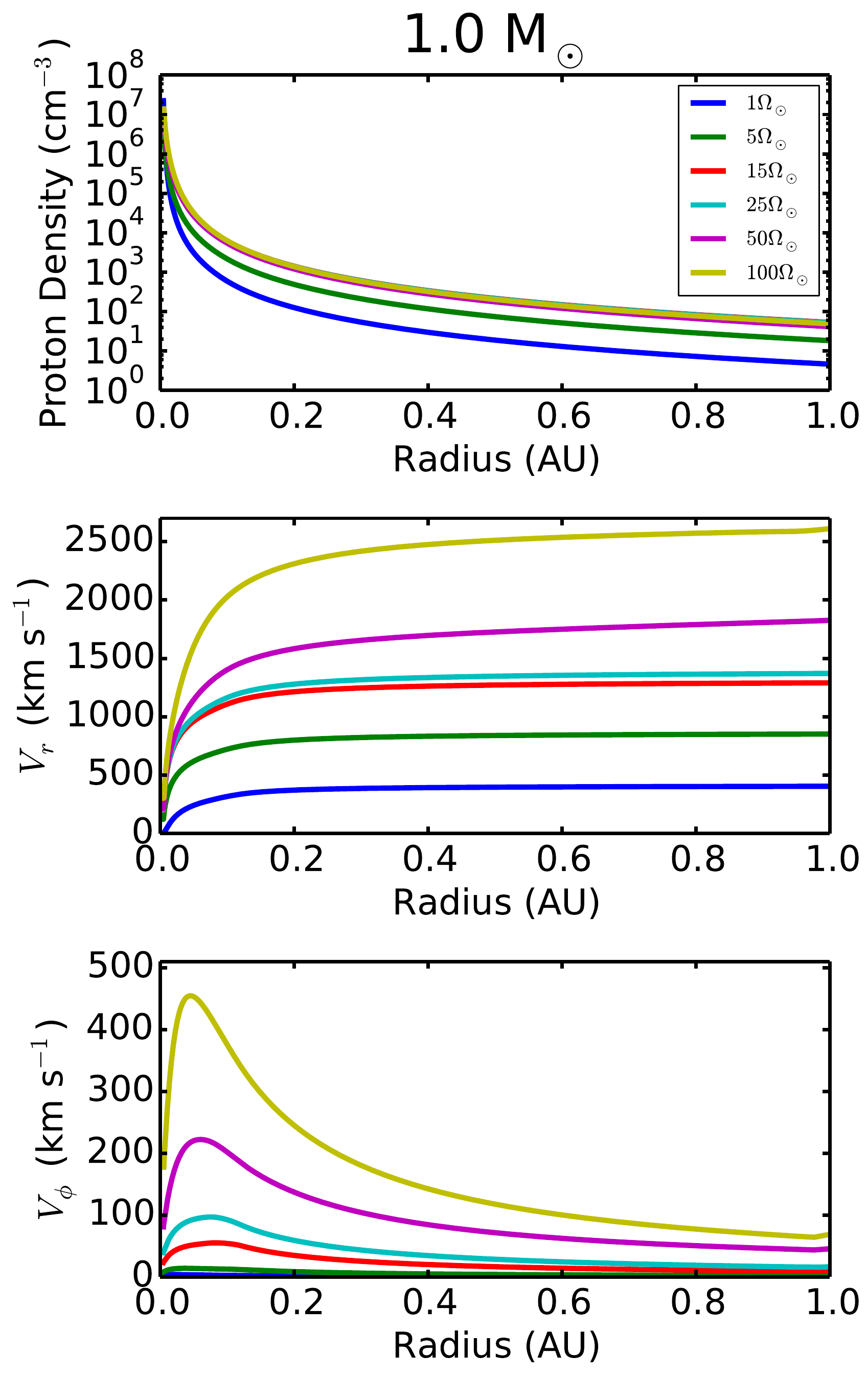}
\caption{
Figure showing the wind properties for stars with masses of 0.5~M$_\odot$ (\emph{left-column}) and 1.0~M$_\odot$ (\emph{right-column}) for different rotation rates.
Note the different scales on the y-axes in the two columns. 
}
\label{fig:WindProperties}
\end{figure*}

Once the wind temperature has been set, to get the base density, we choose the value that gives us the desired mass loss rate.
We estimate a star's mass loss rate using
\begin{equation} \label{eqn:Mdot}
\dot{M}_\star = \dot{M}_\odot \left( \frac{R_\star}{R_\odot} \right)^{2} \left( \frac{\Omega_\star}{\Omega_\odot} \right)^{1.32} \left( \frac{M_\star}{M_\odot} \right)^{-3.36}.
\end{equation}
This relation was derived by \citet{Paper2} by fitting the free parameters in a rotational evolution model to the observational constraints.
A similar relation was derived using similar methods by \citet{Matt15}.
To scale the magnetic field to other stars, we assume that
\begin{equation} \label{eqn:Bstar}
B_\star = B_\odot \left( \frac{\Omega_\star \tau_\star }{\Omega_\odot \tau_\odot } \right)^{1.32},
\end{equation}
where $\tau_\star$ and $\tau_\odot$ are the stellar and solar convective turnover times.
This is based on the analysis of measured stellar global magnetic field strengths by \citet{Vidotto14}.
For the convective turnover times, we use the relation derived by \citet{Wright11}.
In Eqn.~\ref{eqn:Mdot} and Eqn.~\ref{eqn:Bstar}, when $\Omega_star$ is greater than the saturation rotation rate, $\Omega_\mathrm{sat}$, we assume the $\Omega_\star$ dependence saturates, and replace $\Omega_\star$ with $\Omega_\mathrm{sat}$.
The saturation rotation rate is given by
\begin{equation}
\Omega_\mathrm{sat} = 15 \Omega_\odot \left( \frac{\Omega_\star}{\Omega_\odot} \right)^{2.3}.
\end{equation}
This was estimated by \citet{Paper2} from fitting a stellar rotational evolution model to observations. 

Another effect that we take into account is the increase in the equatorial radii of rapid rotators due to centrifugal effects. 
To calculate the equatorial radius, we numerically solve
\begin{equation}
\frac{G M_\star}{R(\theta)} + \frac{1}{2} \Omega_\star^2 R(\theta)^2 \sin^2 \theta = \frac{G M_\star}{R_\mathrm{p}}
\end{equation}
where $R(\theta)$ is the stellar radius at colatitude $\theta$ and $R_p$ is the polar radius (see e.g. Eqn.~2.10 of \citealt{MaederBook09}).
This equation simply states that the effective gravitational potential is uniform over the stellar surface.
For simplicity, we assume \mbox{$R_\mathrm{p} = R_\star$}, i.e. the polar radius is the radius the entire star would have in the absence of rotation\footnotemark. 
Taking the bulging of the star into account is important for the winds of very rapid rotators and is essential for calculating the break up rotation rate.

\footnotetext{
Based on Fig.~2.7 of \citet{MaederBook09}, we expect this assumption to be accurate, with only a few percent difference between $R_\mathrm{p}$ and $R_\star$.
}

\section{Wind temperature: constraints from rotational evolution} \label{sect:ModelAorB}

In this section, we consider the different influences of Model~A and Model~B on the mass loss rates of solar mass stars.
The importance of these models is not that either is likely to be correct, but that they represent different possibilities for how wind acceleration changes with the activity level of the star.
Model~A represents cases in which the winds of more active stars are faster than the winds of less active stars, either because they are hotter or because of stronger acceleration from waves, or a combination of the two. 
An example of this is the model used by \citet{Airapetian16}. 
Model~B represents the case where the wind speeds are constant, and the mass loss rates change only because of changes in the densities. 
Examples of this assumption can be found in \citet{Wood02} and \citet{CranmerSaar11}.

\begin{figure*}
\centering
\includegraphics[trim = 0mm 0mm 0mm 0mm, clip=true,width=0.45\textwidth]{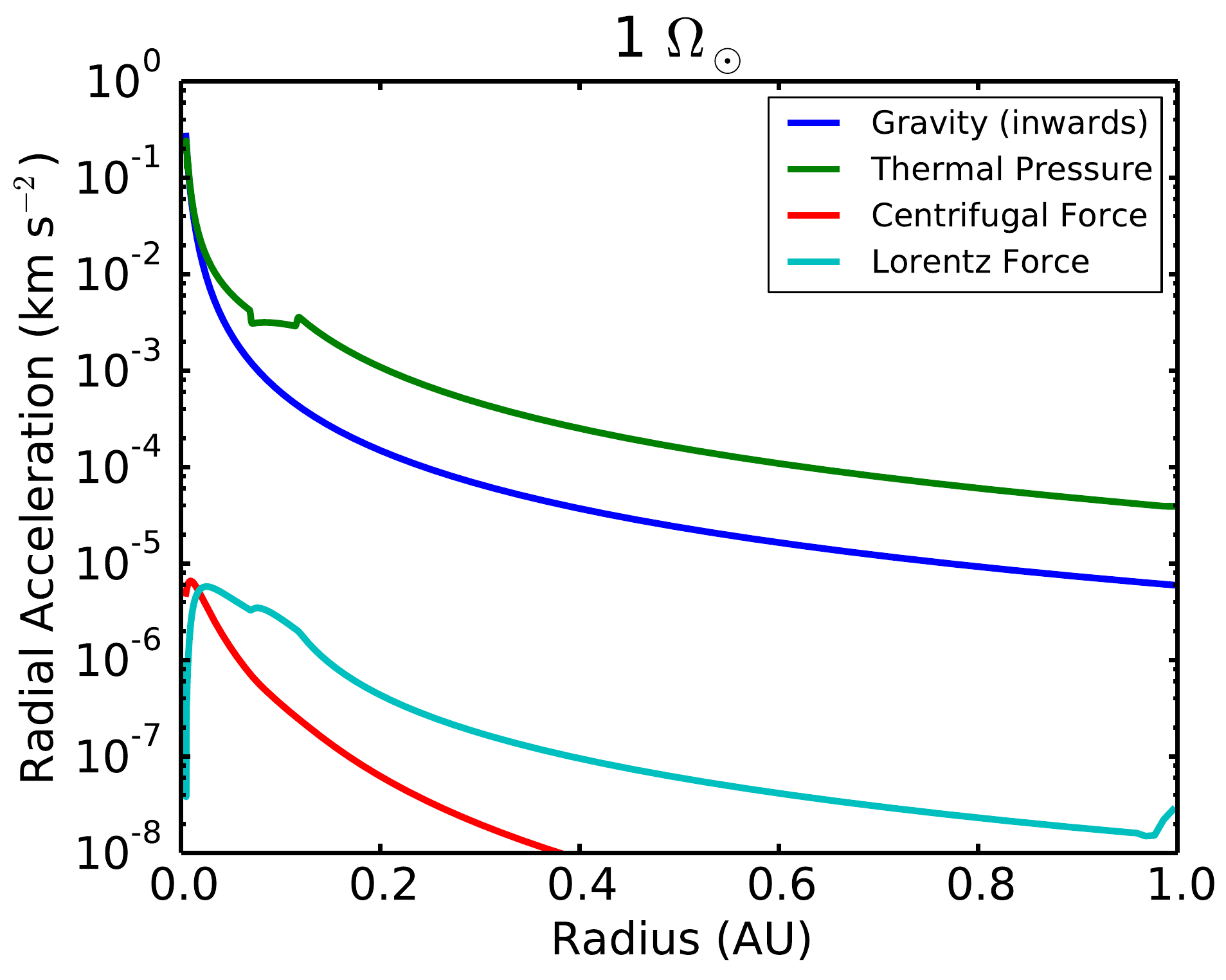}
\includegraphics[trim = 0mm 0mm 0mm 0mm, clip=true,width=0.45\textwidth]{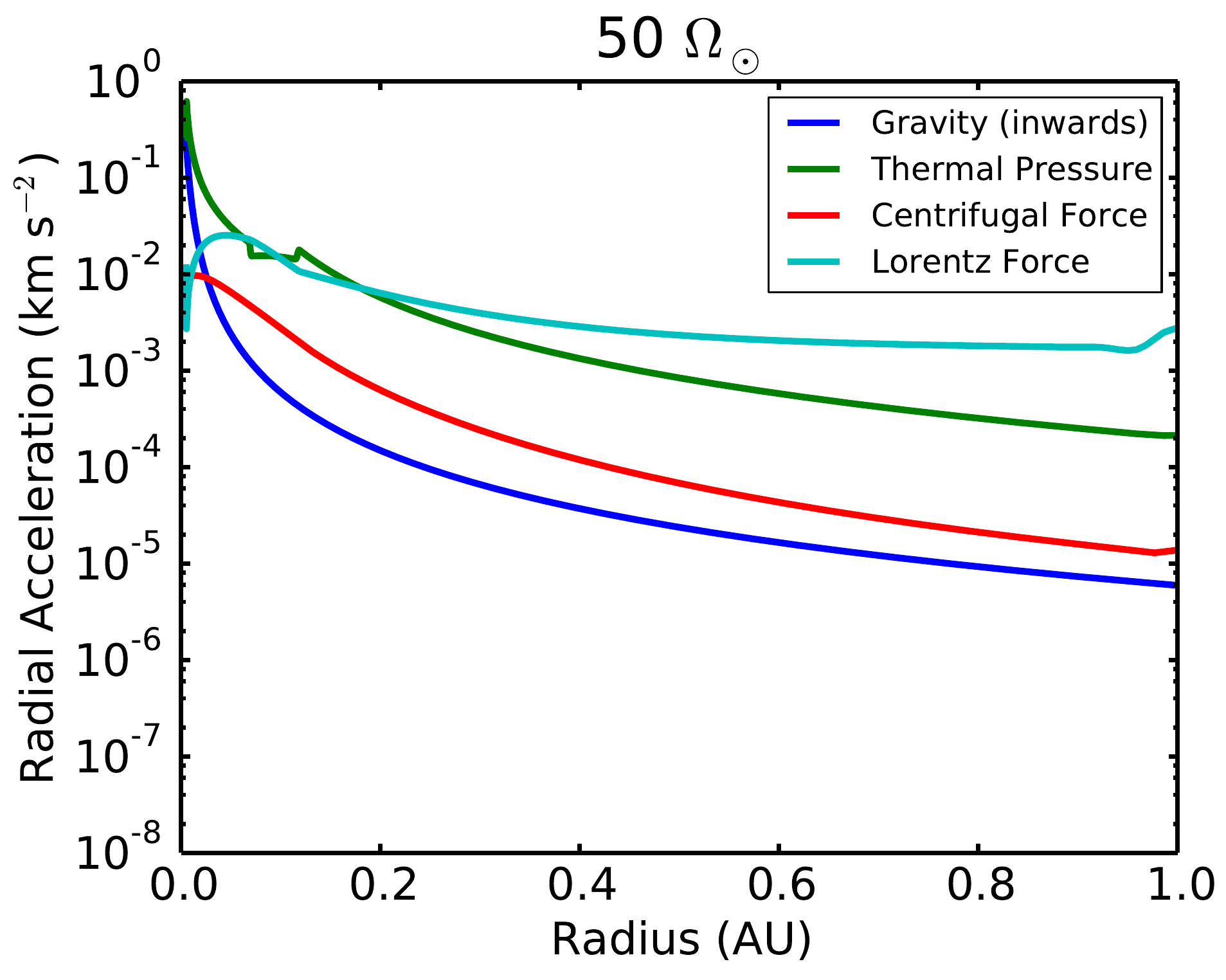}
\caption{
Figure showing the radial variations in the forces that cause acceleration in our wind model; these are the four terms on the RHS of Eqn.~\ref{eqn:accelerations}.
Both cases are for solar mass stars. 
}
\label{fig:radialaccelerations}
\end{figure*}

\begin{figure}
\centering
\includegraphics[trim = 0mm 0mm 0mm 0mm, clip=true,width=0.45\textwidth]{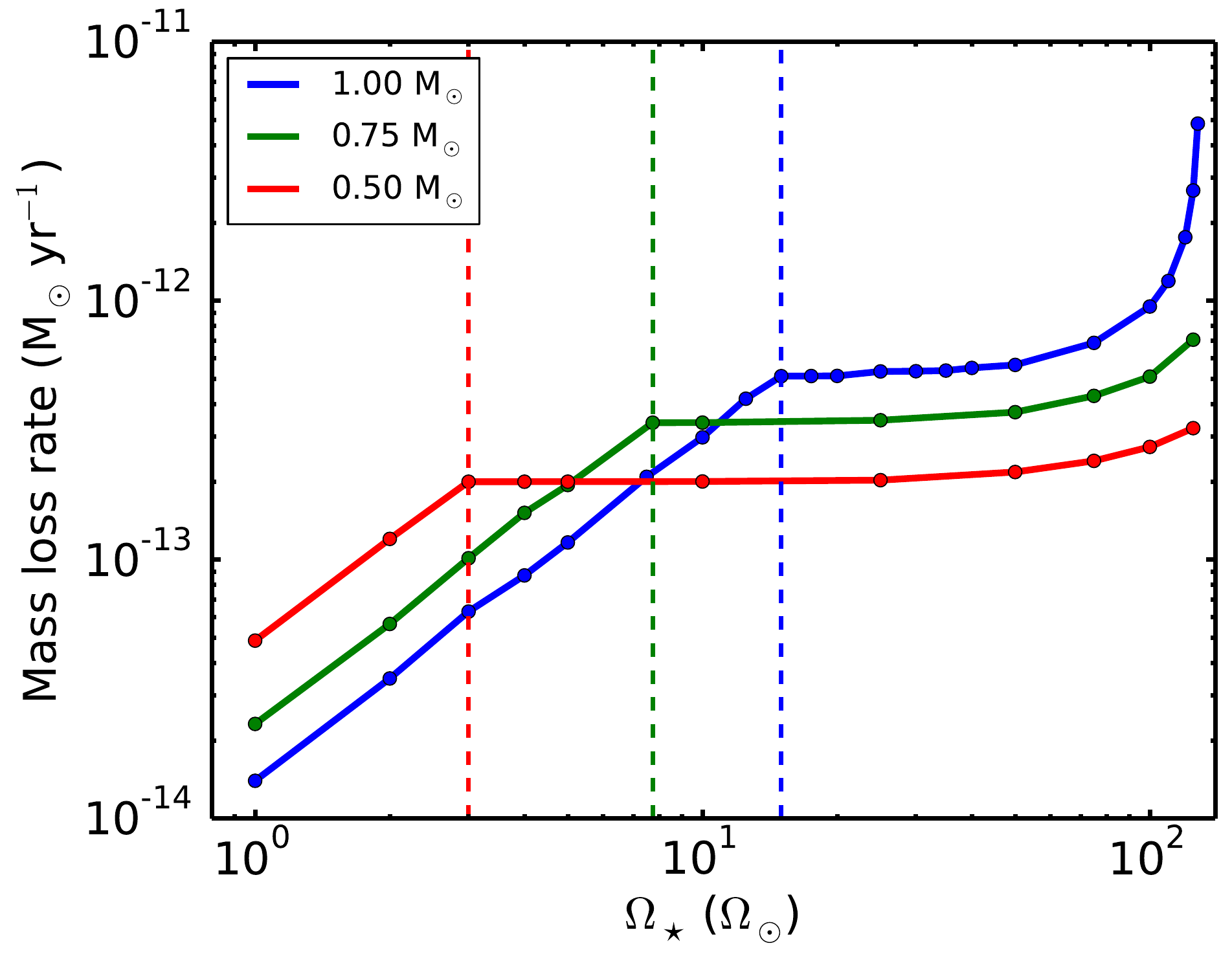}
\includegraphics[trim = 0mm 0mm 0mm 0mm, clip=true,width=0.45\textwidth]{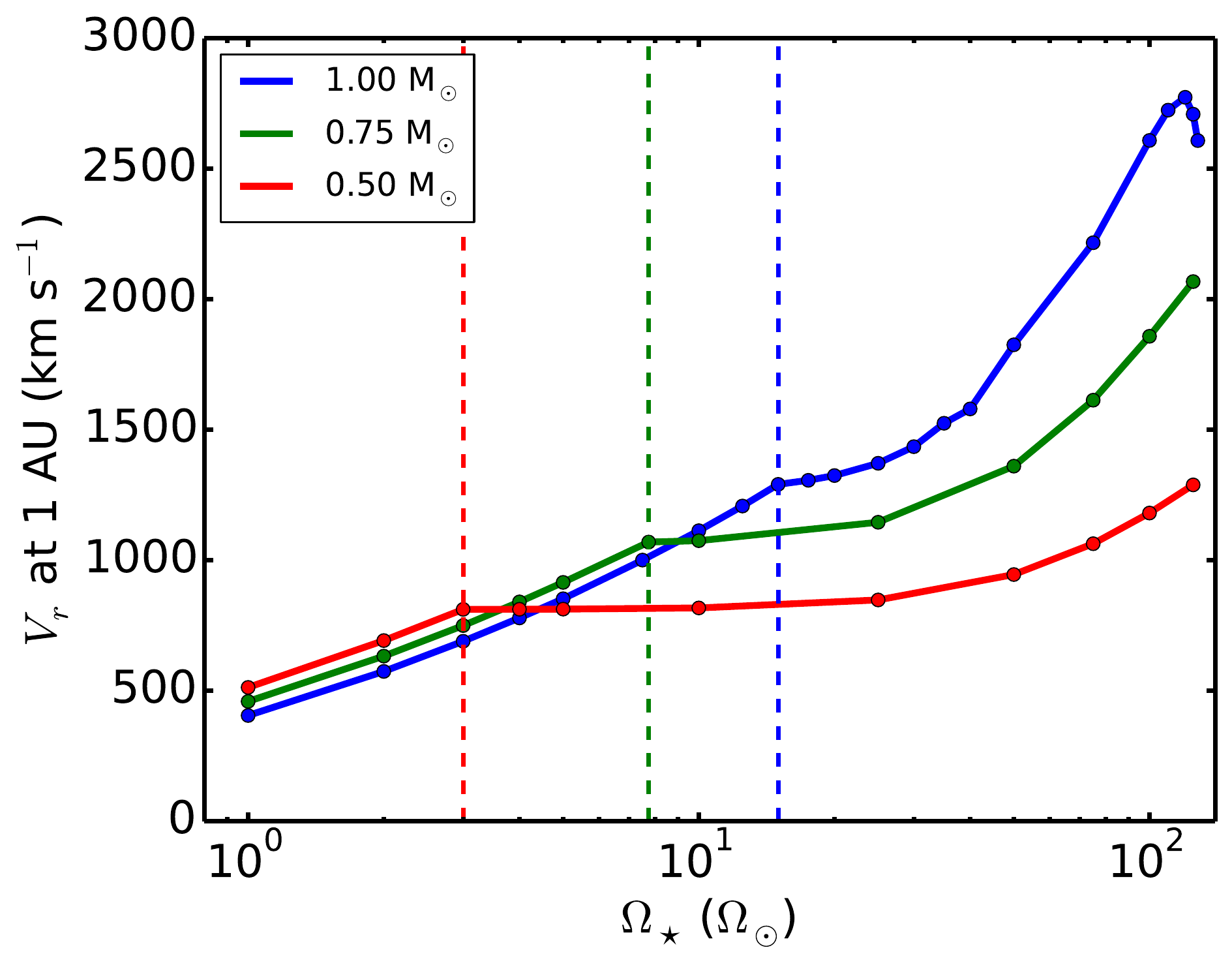}
\caption{
Figure showing the mass loss rate (\emph{upper-panel}) and radial outflow speed (\emph{lower-panel}) as a function of rotation for three stellar masses.
The small circles show the individual simulations and the vertical dashed lines show the saturation thresholds. 
}
\label{fig:MdotVrOmega}
\end{figure}

In Fig.~\ref{fig:MdotsSolar}, we show the dependence of mass loss rate on stellar rotation rate for both Model~A and Model~B. 
In the unsaturated regime, we force the mass loss rate to increase according to \mbox{$\dot{M}_\star \propto \Omega_\star^{1.33}$}, as estimated from observed stellar rotational evolution (\citealt{Paper2}). 
In the saturated regime, we hold constant the base density, temperature, and radial field, and $\dot{M}_\star$ is allowed to change as a result of the changes in the centrifugal and Lorentz forces. 
We therefore have in Model~A a higher base temperature and a lower base density than in Model~B, for all models except for the current solar wind case. 
Both models are by definition identical for the current Sun.

We find that at rapid rotation, Model~A and Model~B lead to significantly different mass loss rates. 
In Model~A, $\dot{M}_\star$ saturates at the saturation threshold ($15\Omega_\odot$) and remains approximately constant until $\sim100\Omega_\odot$ when it then begins to increase rapidly with faster rotation. 
In Model~B, the mass loss rate never saturates and continues to increase with increasing rotation rate at approximately the same rate as in the unsaturated regime. 

This difference is caused by the different acceleration profiles of the winds in the two models and the important fact that the mass loss rates in winds are determined by what happens in the lower parts of the wind where the material is subsonic and subalfv\'{e}nic.
The wind speeds at 1~AU and the radii of the sonic points as functions of the rotation rate are shown in Fig.~\ref{fig:SonicSolar}.
In Model~A, the winds of more rapidly rotating stars accelerate faster and therefore become supersonic closer to the stellar surface. 
The increasing centrifugal and Lorentz forces acting on the winds have a smaller effect on the mass loss rate since they are primarily significant at larger radii.  
In Model~B, since the wind temperatures remain constant, the sonic point remains far from the stellar surface at almost all rotation rates, so the centrifugal and Lorentz forces are significant in the subsonic wind.
The result is that even if the primary driving mechanisms of the winds saturate, $\dot{M}_\star$ still increases quickly with rotation because of magneto-centrifugal acceleration.

It is interesting to consider stellar rotational evolution since spin-down is a direct consequence of stellar winds and is well constrained observationally (\citealt{Bouvier14}). 
We can describe the spin-down rate of a star as \mbox{$d\Omega_\star / dt \propto \Omega_\star^a$}.
In the unsaturated regime, \mbox{$a \approx 3$}, and in the saturated regime, \mbox{$a \approx 1$} (\citealt{Matt15}).
The wind torque law found by \citet{Matt12} approximately implies that \mbox{$d\Omega_\star / dt \propto B_\star^{0.87} \dot{M}^{0.56} \Omega_\star$}.
Inserting the $\Omega_\star$ dependences of $\dot{M}_\star$ from Eqn.~\ref{eqn:Mdot} and $B_\star$ from Eqn.~\ref{eqn:Bstar} gives \mbox{$a \approx 2.89$}, as expected for the unsaturated regime.
In the saturated regime, we instead get \mbox{$a \approx 1$} if both $\dot{M}_\star$ and $B_\star$ saturate, and \mbox{$a \approx 1.74$} if only $B_\star$ saturates, suggesting that $\dot{M}_\star$ saturation is observationally necessary.
This is further illustrated in Fig.~\ref{fig:RotEvo} where we show the rotational evolution of a solar mass star with a 100~Myr rotation rate of $70\Omega_\odot$.
This corresponds to the 90th percentile of the rotation distribution at this age.
We use the rotational evolution model described in \citet{Paper2} to evolve the star's rotation until 1~Gyr assuming saturation of $\dot{M}_\star$ (blue line) and no saturation of $\dot{M}_\star$ (green line). 
The red triangles show the observed 90th percentile at different ages, which are clearly better fitted by the blue line.
The necessity of $\dot{M}_\star$ saturation would likely be even clearer if we considered the rotational evolution of rapid rotators before 100 Myr.
We can conclude from these considerations that the $\dot{M}_\star$--$\Omega_\star$ relation shown in Fig.~\ref{fig:MdotsSolar} for Model~B is not consistent with the observed rotational evolution of rapid rotators

\section{Results: Stellar winds in the FMR regime} \label{sect:FMRwinds}

We now consider only Model~A for calculating wind temperatures. 
In Fig.~\ref{fig:WindProperties}, we show the wind densities, and radial and azimuthal speeds for 0.5~M$_\odot$ and 1.0~M$_\odot$ stars with a wide range of rotation rates. 

In the 1.0~M$_\odot$ case, the wind has the acceleration profile of the current solar wind, with a radial outflow velocity at 1~AU of 400~km~s$^{-1}$ and an azimuthal velocity of 0.5~km~s$^{-1}$.
Going to faster rotation, the winds accelerate to higher speeds due to their higher temperatures, with a speed of about 1300~km~s$^{-1}$ for stars rotating at $15\Omega_\odot$. 
This is entirely due to increased thermal acceleration.
At this point, the wind driving saturates and the radial outflow speed depends only weakly on rotation until $\sim 50\Omega_\odot$, where magneto-centrifugal wind acceleration becomes significant. 
At more rapid rotation, the wind speed depends strongly on $\Omega_\star$.   
At $50\Omega_\odot$, the wind has a $V_r$ at 1~AU of 1800~km~s$^{-1}$, and at $100\Omega_\odot$, this value is 2600~km~s$^{-1}$
In all cases, the azimuthal component of the wind velocity, which is strongly $\Omega_\star$-dependent, peaks close to the star and then decreases to insignificant values by 1~AU.

The acceleration of the solution is described by 
\begin{equation} \label{eqn:accelerations}
V_r \frac{d V_r}{dr} = - \frac{1}{\rho} \frac{dp}{dr} - \frac{G M_\star}{r^2} + \frac{V_\phi^2}{r} - \frac{B_\phi}{4\pi \rho r} \frac{d}{dr} \left( r B_\phi \right).
\end{equation}
This is Eqn.~9.18 of \citet{LamersCassinelli99}.
The terms on the RHS correspond to each of the forces that influence the radial acceleration of the wind; from left to right, they are the thermal pressure gradient, gravity, the centrifugal force, and the radial component of the Lorentz force. 
In the absence of stellar rotation and/or the magnetic field, the final two terms are zero and the equation corresponds to the model of \citet{Parker58}.
We show in Fig.~\ref{fig:radialaccelerations} each of these terms as a function of radius for rotation rates of $1\Omega_\star$ and $50\Omega_\star$.  
For the $1\Omega_\star$ case, the centrifugal and Lorentz force terms are negligible at all radii. 
For the $50\Omega_\star$ case, the centrifugal term is negligible, but the Lorentz force is stronger than the thermal pressure term beyond 0.2~AU.
Since the wind is already supersonic at 0.2~AU, $\dot{M}_\star$ is still determined by the thermal acceleration.

The wind acceleration in the 0.5~M$_\odot$ case, shown in Fig.~\ref{fig:WindProperties}, leads to much slower winds for the rapid rotators. 
The 1~AU wind speed for the 100$\Omega_\odot$ case is 1200~km~s$^{-1}$, which is half that of the 1.0~M$_\odot$ case. 
To understand this, consider the Michel velocity (\citealt{Michel69}) which can be used as an estimate of the wind outflow speed far from the star in the FMR regime (\citealt{BelcherMacGregor76}).
The Michel velocity is given by
\begin{equation}
V_\mathrm{M} = \left( \frac{r^4 B_r^2 \Omega_\star^2}{\dot{M}_\star} \right)^{\frac{1}{3}}.
\end{equation}
Since \mbox{$B_r \propto r^{-2}$}, this is a constant throughout the wind. 
At rapid rotation, due to the lower saturation threshold, $\dot{M}_\star$ and $B_\star$ are lower for lower mass stars in our model. 
These two almost cancel each other out, leading to only a small effect on $V_\mathrm{M}$.
The main reason for the lower wind speeds of lower mass stars is that $B_r$ at large radii is lower, due not to the weaker field at the stellar surface, but due to the smaller stellar radius, since \mbox{$B_r(r) = B_\star \left( r / R_\star \right)^{-2}$}.
The dependences of $\dot{M}_\star$ and $V_r$ at 1~AU on $\Omega_\star$ can be clearly seen in Fig.~\ref{fig:MdotVrOmega}.

\section{Conclusions} \label{sect:conclusions}

In this paper, we study the winds of rapidly rotating stars using recent knowledge of stellar magnetic fields. 
We show that solar mass stars rotating at 100$\Omega_\odot$ have winds with speeds of $\sim 2500$~km~s$^{-1}$.
Such rapid rotators have be seen in young stellar clusters (e.g. \citealt{Hartman10}), and it is possible that the Sun was once rotating at such a rate. 
Even greater wind speeds should be expected given that we use only average magnetic field strengths, and in reality, the global magnetic field of a star will vary over a large range. 
We should expect that as the global fields of stars vary, both stochastically and cyclically, the winds speeds will vary in response. 
Given the influence that stellar winds can have one the atmospheric evolution of planetary atmospheres, understanding the properties of stellar winds in the FMR regime can be important for our understanding of the atmospheric evolution of the solar system terrestrial planets. 

Based on the solar relation between flares and coronal mass ejections (CMEs), and the high flare activity of active stars, it has been suggested that the winds of active stars are dominated by CMEs (\citealt{Aarnio12}; \citealt{Drake13}), though their actual existence is still unclear (\citealt{Leitzinger14}).
The influences on highly time-dependent CME winds of the magneto-centrifugal forces of rapid rotators is currently unexplored.

Probably the most important problem in the study of stellar winds of low mass stars is the lack of direct observational constraints on wind properties. 
Indirect methods include measuring astrospheric Ly$\alpha$ absorption (\citealt{Wood14}), getting upper limits on radio emission from winds (\citealt{Gaidos00}), fitting stellar rotational evolution (\citealt{Matt15}), and the analysis of planetary transits (\citealt{Kislyakova14}).
The most important result in this paper is the suggestion that assumptions for the wind speeds of rapid rotators can be tested using rotational evolution.
If we assume that wind speeds are the same for stars with different activity levels, we find almost no saturation of the mass loss rate at fast rotation. 
The lack of saturation is likely inconsistent with observed rotational evolution, meaning that this could be an indirect observational constraint on stellar wind properties. 
Specifically, the winds of more rapidly rotating stars need to be faster so that they become supersonic closer to the stellar surface; if this does not happen, magneto-centrifugal acceleration can significantly influence the mass loss rate, causing the most rapidly rotating stars to spin down faster than is observed. 

This argument needs to be explored with more complex and internally consistent models before it can be made with any certainty. 
The 1D equatorial model that we use here will overestimate the increase in $\dot{M}_\star$ due to magneto-centrifugal effects, which should be fixed by applying 2D and 3D MHD models. 
Such models will also be able to take into account the meridional components of the velocity, and meridional gradients in all quantities (i.e. $\partial/\partial \theta$ terms), which are ignored in the Weber-Davis model (\citealt{SuessNerney73}).
Given the lack of observational constraints on the winds of low mass stars, if these results can be confirmed by more sophisticated theoretical work, it could provide an important constraint on future stellar wind models.


\section{Acknowledgments}

CPJ acknowledges the support of the FWF NFN project S116601-N16 ``Pathways to Habitability: From Disks to Active Stars, Planets and Life'', and the related FWF NFN subproject S116604-N16. 
This publication is supported by the Austrian Science Fund (FWF).


\appendix

\section{Solving the Weber-Davis (WD) model} \label{appendix:analyticWD}

To solve the isothermal WD model, we use the method described by \citet{Preusse05}.
The six input parameters are the wind temperature, $T_\mathrm{wind}$, the radial field at the star, $B_{r,\star}$, and the stellar mass, $M_\star$, radius, $R_\star$, and rotation rate, $\Omega_\star$.
From these parameters, the radial structures of the wind density, velocity, and magnetic field can be derived.

Unlike the wind model of \citet{Parker58}, which has one critical radius at the sonic point, the WD model has three critical points. 
This can be seen from Eqn.~9.73 of \citet{LamersCassinelli99} for the radial gradient in the radial component of the wind velocity, $V_r$:
\begin{equation} \label{eqn:dVrdr}
\frac{r}{V_r} \frac{d V_r}{dr} = \frac{ \left( V_r^2 - A_r^2 \right) \left( 2 c_s^2 + V_\phi^2 - \frac{G M_\star}{r} \right) + 2 V_r V_\phi A_r A_\phi }{ \left( V_r^2 - A_r^2 \right) \left( V_r^2 - c_s^2 \right) - V_r^2 A_\phi^2 },
\end{equation} 
where $V_\phi$ is the azimuthal component of the wind velocity in the non-rotating frame of reference, \mbox{$A_r=B_r/\sqrt{4 \pi \rho}$} and \mbox{$A_\phi=B_\phi/\sqrt{4 \pi \rho}$} are the radial and azimuthal components of the Alfv\'{e}n velocity at a given $r$, and $c_s$ is the sound speed.
At a given radius, $V_\phi$ and $A_\phi$ are given by
\begin{equation} \label{eqn:VphiBphi}
V_\phi = \frac{ \frac{V_r^2 \mathscr{L}}{r^2 \Omega_\star} - A_r^2 }{ V_r^2 - A_r^2 },
\hspace{5mm}
B_\phi = B_r \frac{ V_\phi - \Omega_\star r }{ V_r },
\end{equation}
where $\mathscr{L}$ is defined below. 
The denominator of the RHS of Eqn.~\ref{eqn:dVrdr} vanishes when 
\begin{equation}
V_r^2 = V_{r,s}^2 = \frac{1}{2} \left[ c_s^2 + |\mathbf{A}|^2 - \sqrt{ (c_s^2 + |\mathbf{A}|^2)^2 - 4 c_s^2 A_r^2 }  \right],
\end{equation}
and when
\begin{equation}
V_r^2 = V_{r,f}^2 = \frac{1}{2} \left[ c_s^2 + |\mathbf{A}|^2 + \sqrt{ (c_s^2 + |\mathbf{A}|^2)^2 - 4 c_s^2 A_r^2 }  \right],
\end{equation}
where \mbox{$|\mathbf{A}|^2 = A_r^2 + A_\phi^2$}.
Specifically, this is where $V_r$ is equal to the radial component of either the slow or the fast magnetosonic wave velocities, given by $V_{r,s}$ and $V_{r,f}$. 
At these points, the numerator and denominator of the RHS of Eqn.~\ref{eqn:dVrdr} simultaneously vanish.
These two points are the slow and fast points.  
The third critical point is the Alfv\'{e}n point, at radius $r_A$, where $V_r = A_r$, which is between the slow and fast points. 

It is convenient to define \mbox{$x=r/r_A$} and \mbox{$u=V_r/V_A$}, where \mbox{$V_A = V_r (r_A) = A_r (r_A)$} is the radial wind speed at the Alfv\'{e}n point, and rewrite the equation for $dV_r/dr$ as
\begin{multline} \label{eqn:dVrdrNormed}
\left[ 1 - \frac{\mathscr{V}}{u^2} + \mathscr{W} \frac{x^2 \left( 1 - x^2 \right)^2 }{(1-x^2 u)^3} \right] x^2 u \frac{du}{dx} 
=  \\
2 \mathscr{V} x - \mathscr{U} +  \mathscr{W} \left[ \frac{x^3 \left( 1 - u \right)^2}{\left(1-x^2u \right)^2} + \frac{2 x^3 u \left( 1 - u \right) \left( 1 - x^2 \right)}{\left(1-x^2u \right)^3} \right]
\end{multline}
where $\mathscr{U}$, $\mathscr{V}$, and $\mathscr{W}$ are constants given by
\begin{equation}
\mathscr{U} = \frac{G M_\star}{r_A V_A^2}, \hspace{5mm} \mathscr{V} = \frac{c_s^2}{V_A^2},\hspace{5mm} \mathscr{W} = \frac{\left(r_A \Omega_\star \right)^2}{V_A^2}.
\end{equation}
These are Eqns.~9.67 and 9.68 of \citet{LamersCassinelli99}.
It is also convenient to define three more constants in the wind representing mass and magnetic fluxes, and the specific angular momentum:
\begin{equation} \label{eqn:windconstants}
\mathscr{F}_m = r^2 \rho V_r = \frac{\dot{M}}{4 \pi}, 
\hspace{2mm}
\mathscr{F}_B = B_r r^2 = B_{r,\star} R_\star^2 , 
\hspace{2mm} 
\mathscr{L} = \Omega_\star r_A^2 .
\end{equation}
As in \citet{Preusse05}, integrating Eqn.~\ref{eqn:dVrdrNormed} gives
\begin{multline} \label{eqn:dVrdrInted}
\mathscr{W} + \frac{1}{2} u^2 + \frac{1}{2} \mathscr{W} \frac{x^2 \left( 1 - u \right)^2}{\left( 1 - x^2 u \right)^2} 
= \\
\mathscr{V} \ln{x^2} + \mathscr{V} \ln{u} + \frac{\mathscr{U}}{x} + \mathscr{W} \frac{x^2 \left( 1 - u \right)}{1-x^2 u} + C
\end{multline}
where $C$ is the constant of integration.
This equation can easily by solved numerically to get $V_r$ at all radii when the values of $C$, $r_A$, and $V_A$ are known. 
As in \citet{Preusse05}, we find these values by searching for the value of $\mathscr{V}$ that gives the correct wind solution.

For each guess of $\mathscr{V}$, we first calculate the corresponding $V_A$ and $r_A$ values from Eqn.~\ref{eqn:windconstants} and
\begin{equation}
r_A = \frac{F_B}{\sqrt{4 \pi F_m V_A}}.
\end{equation}
From these, we get $\mathscr{U}$, $\mathscr{W}$, and $\mathscr{L}$.
Then, we find the values of $r$ and $V_r$ where the nominator and denominator of Eqn.~\ref{eqn:dVrdr} simultaneously vanish.\footnotemark
One of these points is where \mbox{$r<r_A$} and \mbox{$V_r<V_A$} and corresponds to the slow point, and the other is where \mbox{$r>r_A$} and \mbox{$V_r>V_A$} and corresponds to the fast point.
We then use Eqn.~\ref{eqn:dVrdrInted} to calculate $C$ at the slow and fast points.
We start by taking an initial guess of $\mathscr{V}$ and search for the value of $\mathscr{V}$ at which $C$ is the same at the slow and fast points. 
With the correct value of $\mathscr{V}$, we then solve Eqn.~\ref{eqn:dVrdrInted} numerically to get $V_r (r)$, then use $\mathscr{F}_m$ and $\mathscr{F}_B$ to get $\rho (r)$ and $B_r (r)$, and finally Eqn.~\ref{eqn:VphiBphi} to get $V_\phi (r)$ and $B_\phi (r)$.

\footnotetext{For this, we use the \mbox{\emph{optimize.root()}} function of the SciPy package (\citealt{SciPyWebsite}).
For finding the slow point, \mbox{\emph{optimize.root()}} required a good initial guess, which we make by eye.}


\bibliographystyle{aa}
\bibliography{mybib}

\end{document}